\documentclass[11pt]{article}
\usepackage[utf8x]{inputenc}
\usepackage[left=2.5cm,right=2.5cm,top=2cm,bottom=2cm]{geometry}

\usepackage{color}
\usepackage{graphicx}
\usepackage{amsfonts}
\usepackage{amsmath}
\usepackage[colorlinks=true,pdfborder={0 0 0},linkcolor=blue,citecolor=violet]{hyperref}
\usepackage{hyperref,xcolor}
\definecolor{winered}{rgb}{0.5,0,0}

\begin{document}
\def\sigmaB{\sigma}
\def\sigmaBH{\hat{\sigma}}
\def\delR{\delta^\text{regul}_{\Delta C}}
\def\pUp{p^\text{stop}_\text{up}}
\def\EB{{\bar{E}}}
\def\FB{{\bar{F}}}
\def\th{{\hat{t}}}

\title{Solving non-linear integral equations for laser pulse retrieval with Newton's method} 
\author{Michael Jasiulek\footnote{Max-Born-Institut f\"ur Nichtlineare Optik, Max-Born-Stra{\ss}e 2A, 
12489 Berlin, Germany, \textit{jasiulek@mbi-berlin.de}}}

\maketitle

\begin{abstract}

We present an algorithm based on numerical techniques that have become standard for
solving nonlinear integral equations: Newton’s method, homotopy continuation, the
multilevel method and random projection to solve the inversion problem that appears when
retrieving the electric field of an ultrashort laser pulse from a 2-dimensional intensity
map measured with Frequency-resolved optical gating (FROG), dispersion-scan or
amplitude-swing experiments. Here we apply the solver to FROG and specify the
necessary modifications for similar integrals. 
Unlike other approaches we transform the integral and work in time-domain where the integral
can be discretised as an over-determined polynomial system and evaluated through list 
auto-correlations.
The solution curve is partially continues and partially 
stochastic, consisting of small linked path segments and enables the computation of optimal 
solutions in the presents of noise. 
Interestingly, this is a novel method to find real solutions of polynomial systems which
are notoriously difficult to find.
%
%
We show how to implement adaptive Tikhonov-type regularization to smooth the solution 
when dealing with noisy data, we compare the results for noisy test data with a least-squares
solver and propose the L-curve method to fine-tune the regularization parameter.
\\
\\
{\small \textit{Keywords---}} nonlinear integral equations, laser pulse retrieval, polynomial root-finding, 
real roots, stochastic optimization, regularization, homotopy continuation


\end{abstract}

\section{Introduction}

In many areas of modern short pulse spectroscopy, attophysics and laser optimization
knowledge of the precise duration and shape of ultrashort laser pulses is elementary. A
direct measurement using electrical detectors is impossible due to their relatively slow
response time and variate pulse retrieval schemes exploiting nonlinear optical effects
have been developed \cite{Walmsley:2009,Geib:2019}.
A common self-referenced technique to measure phase and amplitude of an ultrashort laser
pulse is Frequency-resolved optical gating (FROG)~\cite{Trebino:1993}. A pulse to be
investigated is split in two replicas and then a relative delay is imposed upon the two
which are guided into a nonlinear medium where second harmonic generation (SHG) takes
place. The upconverted light is measured with a spectrometer for varying delays. This
2-dimensional intensity map, the \emph{trace} or spectrogram, encodes all information to
retrieve the enveloping electric field of the investigated pulse. Other modern
self-referenced pulse retrieval schemes varying the amount of dispersion in the optical
path \cite{Miranda:2012} (dispersion-scan) or the relative amplitudes of two pulse
replicas \cite{Alonso:2020} (amplitude-swing) instead of the delay to obtain a
spectrogram.  To invert the associated auto-correlation-like nonlinear integral the most
successful solvers implement the least-squares method using generic optimization or search
methods \cite{Hyyti:2017,Geib:2019}. Although, the related optimization problem is known
to be non-convex these solvers are favorable to approaches inspired by the
Gerchberg-Saxton algorithm~\cite{Gerchberg:1971}
like~\cite{DeLong:1994,Kane:1998real,Sidorenko:2016}, which were the first solvers to be
used. They have the tendency to stagnate, in particular, in the presents of noise which
has been shown in a recent comparative study \cite{Geib:2019}. 

In this paper we present an algorithm linked to a time-domain representation of the integral and based on 
yet unexplored (for the present problem) numerical methods~\cite{Kelley:2018numerical,Kelley:2003solving,Deuflhard:2011} 
that have been successfully applied to other difficult to solve nonlinear integral equations in physics like the Ornstein-Zernike 
equations~\cite{Ornstein:1914accidental,Kelley:2004}, describing the direct correlation functions of molecules in liquids 
and the Chandrasekhar H-equation~\cite{Chandrasekhar:1960radiative} arising in radiative transfer theory, naming two classical
examples. First of all this is Newton's method. Modern Jacobian-free Newton-Krylov methods~\cite{Knoll:2004jacobian}, 
variants of Newton's method, are the basis of large-scale nonlinear solvers like KINSOL, NOX, SNES~\cite{Collier:2020,Heroux:2005,Balay:2012}.
Secondly, homotopy continuation~\cite{Allgower:1993,Morgan:2009,Sommese:2005}, a technique to globalize Newton's method, 
has proven to be reliable and efficient for computing all isolated solutions of polynomial systems and is the primary 
computational method for polynomial solvers like Bertini and PHCpack~\cite{Bates:2013,Verschelde:2011}. 
We employ this method here as there are strong guarantees that a global solution can be found without stagnation
beginning at arbitrary initial data; a main obstacle for the above mentioned solvers which seek the global minimum 
of a non-convex optimization problem.
 
The way the continuation method is implemented here has similarities with techniques in stochastic optimization 
\cite{Pilanci:2017newton,Berahas:2020,Martinsson:2020}. While path tracking towards the 
solution we frequently alternate the random matrices which would be in any case necessary to reduce
the over-determined polynomial system to square form through random projection. 
For each fixed random matrix there is a corresponding homotopy and continuation path, such that
the full solution path is partially continuous and partially stochastic, consisting
of small linked path segments on each one the error decreases monotonically by construction.
This is, up to our knowledge, a novel method for real root finding of polynomial systems \cite{Pan:2011} and 
could be used to find real solutions of similar polynomial systems and optimal solutions, if
noise is present.
 
For the retrieval with realistic noisy experimental data these methods alone would not be sufficient 
because Newton's method has certain smoothness assumptions. For that purpose we chose 
an integral discretization based on grid cell (pixel) surface averages and Tikhonov-type regularization.  
The Tikhonov factor is adaptively decreased during the solution process to obtain a near optimal amount of regularisation
at the solution which can be refined using the L-curve method \cite{Hansen:1999}.
The coarsening capability enables fast computations of low resolution approximants, noise suppression 
and on a hierarchic of finer pixelizations high accuracy retrievals. Multilevel approaches for FROG have 
also been used in \cite{Jafari:2018}.

The list of contents is the following: 
Sec.\ \ref{sec:intro} Notation, integral representation, discretization. 
Sec.\ \ref{sec:setting_poly} Setting real polynomial system, real roots, gauge condition. 
Sec.\ \ref{sec:solver} Polynomial solver, squaring the system, Newton's method, homotopy continuation.
Sec.\ \ref{sec:regul} Adaptive regularizaton for noisy traces.
Sec.\ \ref{sec:application} Application examples, convergence, practical concerns, L-curve method.
Sec.\ \ref{sec:conclusion} Conclusion. 
Appendix\ \ref{app:A} Modifications for similar integrals.
Appendix\ \ref{app:B} Higher order polynomials, splines. 


\section{Notation, integral representation, discretisation} \label{sec:intro}

\begin{figure}[!t]
\centering
\includegraphics[width=0.8\textwidth]{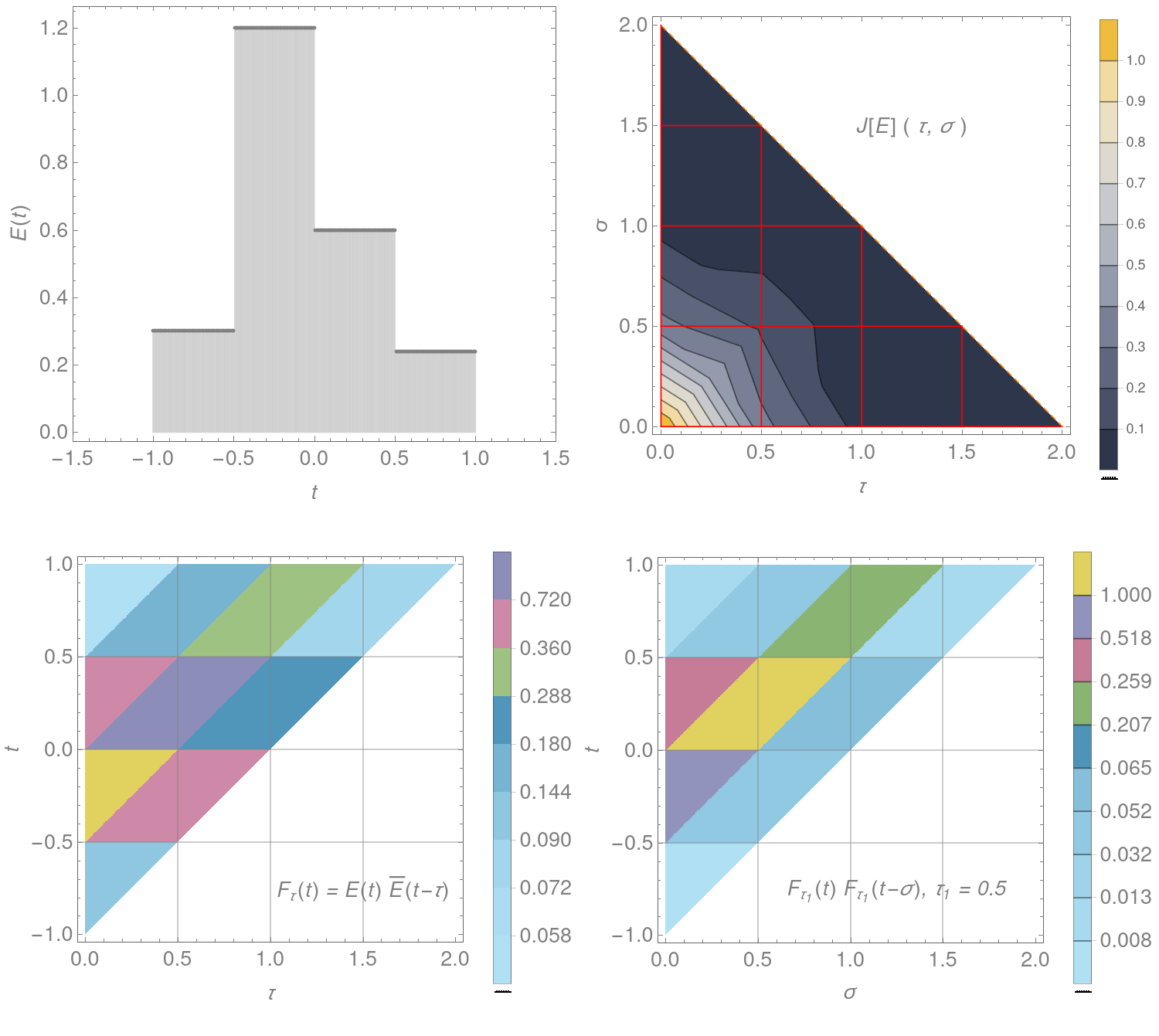}
\caption{\footnotesize Illustrating example: $E(t)$ discretised with a piecewise constant function on $N=4$ intervals (top left).
Then, the associated products $F_{\tau_i}(t)$ and $F_{\tau_i}(t)F_{\tau_i}(t-\sigmaB)$ (bottom) are piecewise constant as well
on small parallelograms such that the nonlinear integral $J[E](\tau,\sigmaB)$ (top right) can be computed via list auto-correlations
along all grid segments (red lines). We consider $J$ only in quadrand $(+,+)$, $\tau,\sigmaB\in[0,2]$ as the other quadrands 
are linked through discrete symmetries. $J$ is non zero only below the diagonal (dashed line) as $E(t)$ is non zero on a
bounded domain $t\in[-1,1]$ by definition. $E(t)$ is generally complex and normalised such that $\text{max}\,|J[E](\tau,\sigmaB)| = 1$.}
\label{fig:exampleN4}
\end{figure}

The nonlinear integral for SHG-FROG is defined as 
\begin{equation} \label{eq:one}
    I[E](\omega,\tau) := \left\lvert \int_{-\infty}^{+\infty} E(t) E(t-\tau) e^{-i\omega t} \,\text{d}t \right\rvert^2,
\end{equation}
where $E(t)$ is a complex function, the enveloping electric field (pulse shape) of the pulse which we 
assume to be non-zero on the interval $t\in[-1,1]$ and zero elsewhere \footnotemark, with time units 
such that this interval has length 2.
The outcome of the FROG experiment is the FROG \textit{trace} 
$I_\text{exp}(\omega,\tau) \approx I[E_\text{in}](\omega,\tau)$ of the pulse to 
be investigated $E_\text{in}(t)$. We obtain $E_\text{in}(t)$ by solving the integral equation
\footnotetext{Setting $E(t)$ on a bounded domain enables clipping of long low-amplitude wings / zooming to the
region of interest on the trace, saving computational cost. A bounded domain can be used as the formalism
is throughout working in time domain.}
\begin{equation} \label{eq:shg_frog}
   I[E](\omega,\tau) - I_\text{exp}(\omega,\tau) = 0. 
\end{equation}
We bring (\ref{eq:one}) into a form better suited for polynomial approximation (removing the 
explicit $t,\omega$ dependence) by Fourier transform\footnotemark $\omega \rightarrow \sigmaB$
\footnotetext{An analogous formulation for the retrieval in frequency domain is possible, see Appendix \ref{app:A}.}
\begin{eqnarray} \label{eq:Jshg}
   J[E](\tau,\sigmaB) &=& 
   \int_{-\infty}^{+\infty} \left( \int_{-\infty}^{+\infty} E(t) E(t-\tau) e^{-i\omega t} \,\text{d}t 
                            \, \int_{-\infty}^{+\infty} \EB(s) \EB(s-\tau) e^{ i\omega s} \,\text{d}s \right)
 e^{i\omega \sigmaB} \,\text{d}\omega \,/\, ( 2\pi) \nonumber \\
 J[E](\tau,\sigmaB) &=& \int_{-\infty}^{+\infty} E(t) E(t-\tau) \EB(t-\sigmaB) \EB(t-\tau-\sigmaB) \,\text{d}t   
\end{eqnarray}
In the first line we have split the absolute value in (\ref{eq:one}) into a complex integral and
its complex conjugate and then applied the relation $\int_{-\infty}^{+\infty} e^{i\omega (r-(t-\sigmaB))} \text{d}\omega = 2\pi \delta(r-(t-\sigmaB))$,
where we call $J[E](\tau,\sigmaB)$ the \emph{double-delay} representation of the SHG-FROG integral and $\EB(t)$ is the complex conjugate of $E(t)$. 
For $E(t)$ non-zero on $t\in [-1,1]$ the trace $J[E](\tau,\sigmaB)$ is non-zero on $\tau,\sigmaB \in [-2,2]$.
In the following we consider only the first quadrant $\tau,\sigmaB \in [0,2]$, as the others are related through discrete symmetries.
We introduce two new function for clarity of notation
\begin{equation}
    F_\tau(t) := E(t)E(t-\tau),\quad G_\sigmaB(t) := E(t)\EB(t-\sigmaB)
\end{equation}
now
\begin{equation} \label{eq:Jauto}
 J[E](\tau,\sigmaB) = \int_{-\infty}^{+\infty} F_\tau(t) \FB_\tau(t-\sigmaB) \text{d}t\quad = \quad \int_{-\infty}^{+\infty} G_\sigmaB(t) G_\sigmaB(t-\tau) \text{d}t.
\end{equation}
For fixed $\tau=\text{const}$ the integral $J[E](\tau,\sigmaB)$ appears to be a one-dimensional auto-correlation of the
function $F_\tau(t)$. We discretize the electric field with a piecewise constant function  
(polynomial of degree zero)\footnotemark, in the context of numerical integration often called \emph{midpoint rule},
\footnotetext{It is possible to use piecewise linear or, more general polynomials or splines, see Appendix \ref{app:B}.}
\begin{equation} \label{eq:0spline}
    E(t) = 
    \begin{cases} 
          0   & t < -1 \quad \text{or} \quad 1 < t \\
          E_k & t \in [t_k,t_{k+1}], \quad k = 0,\dots, N-1 
    \end{cases}
\end{equation}
on a uniform $t$-grid $t_k = -1 + k\cdot h,\, k=0,\dots,N$ with $N$ intervals, where $h=2/N$ 
is the grid spacing. In the same way we define the grids along $\tau$ and $\sigmaB$: $\tau_i, \sigmaB_i = i\cdot h,\, i=0,\dots,N$.
If the delay is equal to an integer multiple of the grid spacing, thus, $\tau = \tau_i$, the product $F_{\tau_i}(t)$ 
is again piecewise constant, see Fig.\ \ref{fig:exampleN4} (bottom left), which we abbreviate as $^{(i)}F_k = F_{\tau_i}(t_k) = E_k E_{k+i}$. 
Then the integrand of $J[E](\tau_i,\sigmaB)$, see Fig.\ \ref{fig:exampleN4} (bottom right), is also piecewise constant
on small parallelograms\footnote{The integration boundaries depend on $\sigmaBH$, for the upper / lower triangles 
they are $\int_{\sigmaBH-1}^1d\hat{t}$ / $\int_{-1}^{\sigmaBH-1}d\hat{t}$.} and integration over $t$ disassembles into two sums of 
$N$ sub-integrals for the $j$th $\sigmaB$ interval $\sigmaB \in [\sigmaB_j, \sigmaB_{j+1}]$ 
($j$th column in Fig.\ \ref{fig:exampleN4} (bottom right))
\begin{eqnarray} \label{eq:list-auto}
J[E](\tau_i,\sigmaB) &=& h\int_{\sigmaBH-1}^1 d\hat{t}\sum_{k=1}^N\,^{(i)}F_k\, ^{(i)} \FB_{k+j}
                   \,\, + \,\, h\int_{-1}^{\sigmaBH-1} d\hat{t}\sum_{k=1}^N\, ^{(i)}F_k\, ^{(i)} \FB_{k+j+1}, \, \sigmaB \in [\sigmaB_j, \sigmaB_{j+1}], \\
J[E](\tau_i,\sigmaB) &=& h(2 - \sigmaBH)\, \text{corr}( ^{(i)}F_k, {^{(i)} \FB}_{k} )_j + h\,\sigmaBH\, \text{corr}( ^{(i)}F_k, {^{(i)} \FB}_{k} )_{j+1}  
\end{eqnarray}
where $\sigmaBH \in [0,2]$ and $\hat{t}\in[-1,1]$ are local coordinates on the intervals $[\sigmaB_j,\sigmaB_{j+1}]$, $[t_k,t_{k+1}]$.
The first sum is collecting all small upper triangles per column in Fig.\ \ref{fig:exampleN4} (bottom right) and the second the lower triangles.
The expression $\text{corr}( ^{(i)}F_k, {^{(i)} \FB}_{k} )_j := \sum_{k=1}^N\,^{(i)}F_k\, ^{(i)} \FB_{k+j}$, $i = 0,\dots,N-1$ 
denotes the list auto-correlations of $^{(i)}F_k$ that can be computed with complexity $N\cdot (N\,log(N))$ \footnotemark. 
\footnotetext{Alternatively, for the relatively small $N$ consider here, the direct method to compute the
correlation is more efficient for $N<1000$, Subsec.~``FFT versus Direct Convolution'' \cite{SASPWEB:2011},
than the FFT-based variant when parallelised on thousands of cores.}
 
Eq.~(\ref{eq:list-auto}) and the equivalent for $G_\sigmaB(t)$ gives the nonlinear integral along all 
grid segments $[\tau_i,\tau_{i+1}]$, $[\sigmaB_j,\sigmaB_{j+1}],\, i,j = 0,\dots,N-1$, red lines in Fig.~\ref{fig:exampleN4} (top right).
Now $N$ could be chosen such that the $\tau_i$ overlap with the points of the experimental data, assuming an equally-spaced 
grid with $K$ points along $\tau$, and the integral equation be solved similar to what follows. 
As a measured trace is normally noisy, the better way to go is setting up a pixelwise instead of a pointwise representation 
of the equation. Moreover, on a coarse-graining hierarchy of smaller grids $N_1<N_2< \dots < K$ the solver is faster and may
resolve long- and short-wavelength components successively.  

At first, we pixelize the integral $J[E](\tau,\sigmaB)$. For the single pixel with grid coordinates 
$(\tau_i, \sigmaB_j)$ (lower left corner) we linearly interpolate the values
from the left pixel boundary to the right boundary 
\begin{equation} \label{eq:lowerLeft}
    J[E](\hat{\tau}, \sigmaBH)_{ij}^\text{left right} = J[E](\tau_i, \sigmaBH) ( 1 - \hat{\tau}/2 ) + J[E](\tau_{i+1}, \sigmaBH)\, \hat{\tau}/2,
\end{equation}
to have the integral approximated inside the pixel.
As before an over hat denotes local pixel coordinates $\hat{\tau}, \sigmaBH \in [0,2]$. 
Then we integrate $J[E](\hat{\tau}, \sigmaBH)_{ij}^\text{left right}$ over the pixel surface 
normalised by its area to obtain the dimensionless pixel average 
\begin{eqnarray} \label{eq:corrLeftRight}
 \langle J[E]_{ij}^\text{left right} \rangle := \int_0^2 \int_0^2 J[E](\hat{\tau}, \sigmaBH)_{ij}^\text{left right}\,  
 d\hat{\tau} d\sigmaBH \,/ \int_0^2 \int_0^2 \, d\hat{\tau} d\sigmaBH  =  
 \frac{1}{2} h\,\text{\LARGE(}\, \text{corr}( {^{(i  )} F}_k, {^{(i  )} \FB}_{k} )_j     + \\
 \text{corr}( {^{(i+1)}F}_k, {^{(i+1)} \FB}_{k} )_{j} +  
                     \text{corr}( {^{(i+1)} F}_k, {^{(i+1)} \FB}_{k} )_{j+1} + \text{corr}( {^{(i  )}F}_k, {^{(i  )} \FB}_{k} )_{j+1} \,\text{\LARGE)}. \nonumber 
\end{eqnarray}
The analog can be done for the bottom and top boundary and the correlation coefficients \\ 
$\text{corr}( {^{(j)} G}_k, {^{(j)} G}_{k} )_i$ improving the accuracy\footnotemark  of the approximation.
Then the total pixel average is 
\begin{equation} \label{eq:JaveTot}
    \langle J[E]_{ij} \rangle := 
\left( \langle J[E]_{ij}^\text{left right}  \rangle +
       \langle J[E]_{ij}^\text{bottom top}  \rangle \right) / 2.
 \end{equation}
and the pixel average of the nonlinear integral is given by adding up the list correlation coefficients for 
each corner square times $\frac{1}{2} h/2$. 
\footnotetext{For most applications it is enough to set 
$\langle J[E]_{ij} \rangle :=  \langle J[E]_{ij}^\text{left right}  \rangle$ speeding up the
computations by a factor of two, though, sacrificing some accuracy. Note the swapping of indices for the 
coefficients of $G$.\label{footnote:speed}}

At last, the pixel averages of the Fourier transformed measurement trace 
$I_\text{exp}(\omega,\tau) \rightarrow J_\text{exp}(\tau,\sigmaB) \rightarrow \langle J^\text{exp}_{ij} \rangle$ 
have to be computed to setup the polynomial system (\ref{eq:double-delay}) where these values constitute the constant part.
This can be computed using the trapezoidal rule or simply by averaging all data points within a pixel.

\section{Setting real polynomial system, real roots, gauge condition} \label{sec:setting_poly}

The integral equation (\ref{eq:shg_frog}) in double-delay representation is now discretised
\begin{equation} \label{eq:double-delay}
   \langle J[E^+,E^-]_{ij} \rangle - \langle J^\text{exp}_{ij} \rangle = 0,
\end{equation}
as a 4th order polynomial system in the $2N$ (generally complex-valued) new variables $E^+, E^-$ with 
components ${E^+_k, E^-_k}$, 
where we have replaced 
\begin{eqnarray}
   E    &\rightarrow&  E^+ + i\, E^- \quad \text{and}  \\ 
   \EB  &\rightarrow&  E^+ - i\, E^-
\end{eqnarray}
in (\ref{eq:shg_frog}) to get rid of the operation of complex conjugation\footnotemark.
\footnotetext{This step may appear confusing at first sight, as we double the number of variables:
Newton's method requires the nonlinear function to be Lipschitz continuous to guarantee convergence
which the operations of complex conjugation or taking the absolute value prevent, see for example
{1.9.1} in \cite{Kelley:2003solving}. Similar requirements, often overseen, 
come in hand with the gradient descent method when applied to least-squares.} 
 
Note: The so-created polynomial system has, strictly speaking, no exact solution as computing
the pixel averages of the nonlinear integral on the one hand and the pixel averages of the trace
come along with numerical and experimental errors limiting the accuracy. For the polynomial 
solver introduced in Sec. \ref{sec:solver} we employ methods from stochastic optimisation to retrieve 
an optimal solution. 
\begin{figure}[!t]
\centering
\includegraphics[width=1.0\textwidth]{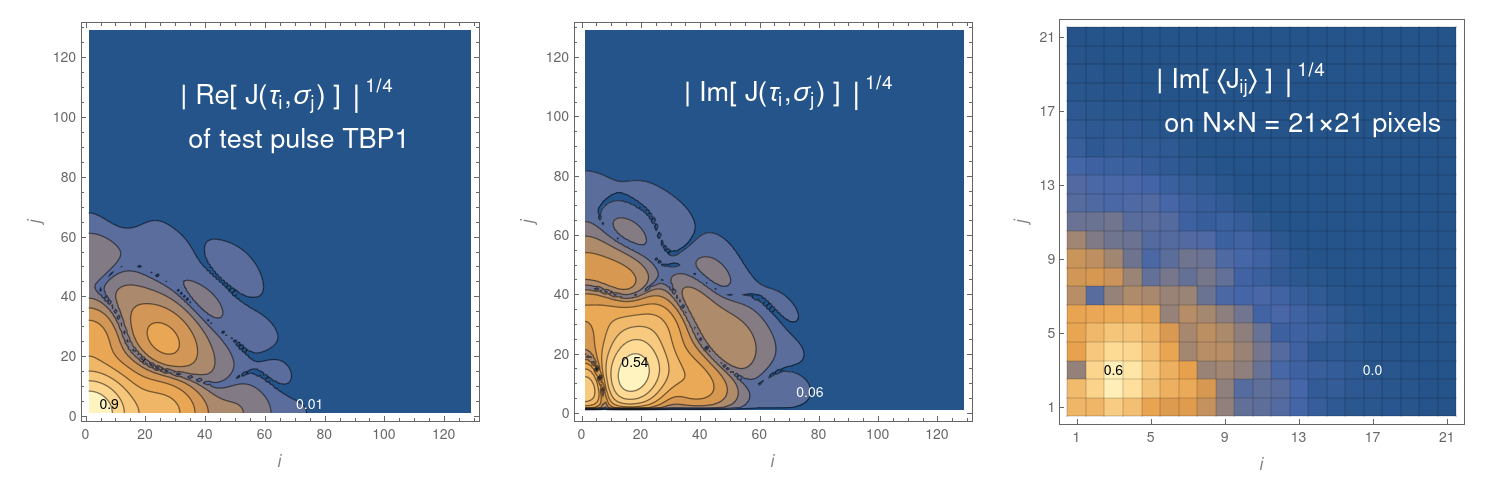}
\caption{\footnotesize Illustrating example: Synthetic measurement trace with $129\times129$ data points 
on $\tau, \sigmaB \in [0,2]$. Coarse-grained data (only imaginary part shown) on $21\times21$ pixels 
(40 data points per pixel) enables fast computation of approximants to initialise refined retrievals.
Every pixel of the lower triangular part ($21\times(21+1)/2$) is associated to one equation in (\ref{eq:ImJ}). 
If $E^+$ and $E^-$ are real roots, then $\widetilde{E} \rightarrow \EB$ and $\langle J[E^+,E^-]_{ij} \rangle^+$ and $\langle J[E^+,E^-]_{ij} \rangle^-$
are real and equivalent to $\text{Re}[ \langle J[E,\EB]_{ij} \rangle]$, $\text{Im}[ \langle J[E,\EB]_{ij} \rangle]$.
Note: The exponent $1/4$ is convenience for data examination as (\ref{eq:double-delay}) constitutes
a 4th order polynomial system in the components $E_k, \widetilde{E}_k$.}
\label{fig:example-trace}
\end{figure}

Clearly, if $E^+, E^-$ are found as real roots of the polynomial system
(\ref{eq:double-delay}), we are dealing with a physical solution. Then $E^+
\rightarrow \text{Re}(E)$, $E^- \rightarrow \text{Im}(E)$ are nothing but the
real and imaginary part of the electric field, though, more generally, complex
solutions exist. 
In the same fashion we create a new polynomial system introducing the linear combinations  
$\langle J_{ij} \rangle^+ = ( \langle J_{ij} \rangle + \langle \bar{J}_{ij} \rangle )/2$ and 
$\langle J_{ij} \rangle^- = ( \langle J_{ij} \rangle - \langle \bar{J}_{ij} \rangle )(-i/2)$
\begin{eqnarray}
   \langle J[E^+,E^-]_{ij} \rangle^+ - \langle J^\text{exp}_{ij} \rangle^+ &=& 0 \label{eq:ReJ} \\ 
   \langle J[E^+,E^-]_{ij} \rangle^- - \langle J^\text{exp}_{ij} \rangle^- &=& 0 \label{eq:ImJ} 
\end{eqnarray}
such that the new system has real coefficients and as long as $E^+$ and $E^-$ are real, eq. (\ref{eq:ReJ}), (\ref{eq:ImJ}) are real
and imaginary part of eq. (\ref{eq:double-delay}). The reasons for these rearrangements are the following:
starting with a real initial iterate, Newton's method remains real and we can stick to real arithmetics,
which is about five times faster than using complex variables, more importantly, we are interested in finding
real roots.
 
For the integral (\ref{eq:Jauto}) the absolute phase as well as the time direction of the electric 
field are not fixed: for any solution $E(t)$, the product $E(t)\cdot \exp( i\, \text{const})$ and $E(-t)$ are also 
solutions. We fix the rotational symmetry by adding the following equation to the system (\ref{eq:ReJ}), (\ref{eq:ImJ}) 
\begin{eqnarray} \label{eq:null}
   \int_{-\infty}^{+\infty} E^+(t) \,\text{d}t -  \int_{-\infty}^{+\infty} E^-(t) \,\text{d}t &=& 0 \quad \Rightarrow \nonumber \\ 
   2/N\, {\sum}_{k=1}^{N} ( E^+_k - E^-_k ) &=& 0 \label{eq:gauge}
\end{eqnarray}
which we call \emph{null gauge condition}\footnotemark, it fixes the absolute complex phase but leaves the overall scaling 
and shape of $E^+(t),\, E^-(t)$ free.
Moreover, its a polynomial equation with real coefficients.
\footnotetext{An alternative null gauge condition with the same properties exists: $\int_{-\infty}^{+\infty} E^+(t) \,\text{d}t = 0$. }

\section{Squaring the system, Newton's method, Homotopy continuation} \label{sec:solver}
The systems (\ref{eq:double-delay}),(\ref{eq:ReJ}),(\ref{eq:ImJ}) consist each of $(N+1)N/2$ equations. 
Only the lower triangular part is non-zero as $E(t)$ is zero beyond the domain $t \in [-1,1]$. 
Such that (\ref{eq:ReJ}),(\ref{eq:ImJ}) contribute $(N+1)N$ equations.
We denote the total system (\ref{eq:ReJ}), (\ref{eq:ImJ}), (\ref{eq:gauge}) with $(N+1)N + 1$ equations as
\begin{equation} \label{eq:FXC}
   F(X) - C_1 = 0, \quad F(X) :=   
   \begin{pmatrix}
      F_1(X_1, \dots, X_{2N}) \\ \vdots \\ F_{(N+1)N+1} (X_{1}, \dots, X_{2N})
   \end{pmatrix}
\end{equation}
where $X = \{ E^+_k, E^-_k \}$ is the list of $2N$ variables and $F(X)$ is the $X$-dependent part of the set of equations 
$F_k(X), k=1,\dots, (N+1)N$ for the lower triangular part of (\ref{eq:double-delay}) flattened to a list and, analogously,
the constant part of (\ref{eq:double-delay}) (pixel trace averages) is flattened to the list
$C_{1\, k}, k=1,\dots, (N+1)N$. The last equation  
$F_{(N+1)N + 1}(X) - C_{1\, (N+1)N + 1} = 0$ is set to be the gauge condition (\ref{eq:gauge}).  
 
\begin{figure}[!t]
\centering
\includegraphics[width=1.0\textwidth]{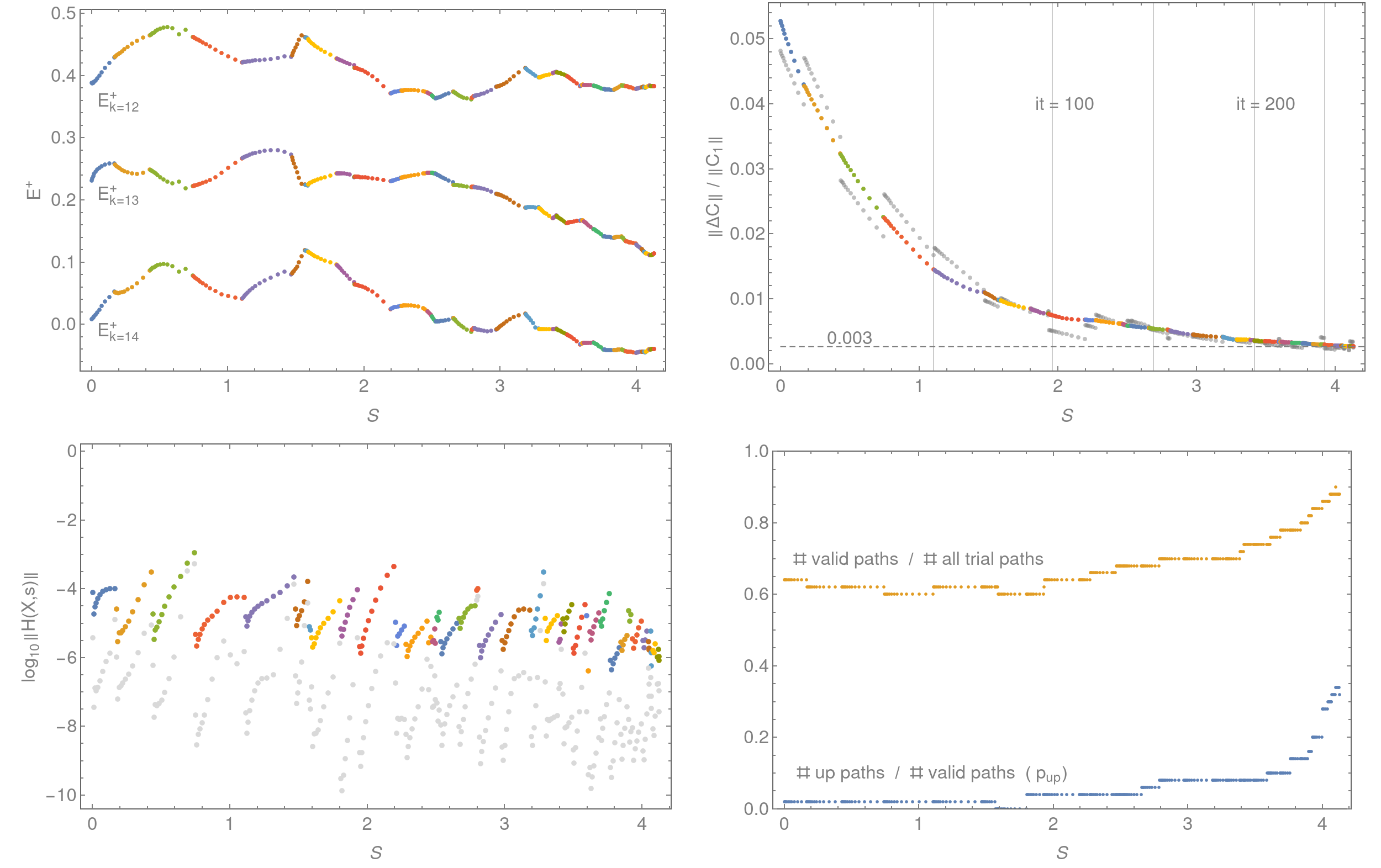}
\caption{\footnotesize Solving polynomial system for $N=15$. Top left: Path tracking the solution curve consisting 
of small path segments each corresponding to a single reduced system with fixed random matrix $M$. Top right: Global distance 
to target trace of full (colored) and reduced (gray) system decreases approxematly exponentially (on each 
segment linearly).
Bottom left: Local residual after predictor step (colored) and after corrector step (Newton step) (gray).
Bottom right: Relative number of up paths (increasing $\lVert \Delta C \rVert$) and valid paths 
(successful Newton step) when doing trial steps at the end of each segment to find a new path (and new $M$)
along which $\lVert \Delta C \rVert$ decreases.}
\label{fig:pathtracker}
\end{figure}

The polynomial system (\ref{eq:FXC}) is overdetermined. Moreover, due to
numerical and experimental errors, it has no exact solution. We
multiply\footnotemark the vector of equations with a random matrix $M$ which
can be reshuffled, having dimensions such that the reduced system has as many
equations as variables. Then, the Jacobian of the reduced system is well
defined and by alternating random matrices stochastic optimisation can be
integrated.
\footnotetext{For Rademacher variables this operation is implemented without any multiplication:
50\% of all equations are added up randomly chosen and the sum of the remaining equations is subtracted to 
obtain one new equation. Moreover, Rademacher variables do not rescale the noise.}
Here we choose $M$ with i.i.d.\ Rademacher random variables (taking values $\{ -1,+1 \}$
with probability $1/2$) to reduce the first $(N+1)N$ equations to $2N-1$ and attach the gauge condition as before 
at the end. 
We denote the reduced system as 
\begin{equation} \label{eq:FXCM}
   F^M(X) - C_1^M = 0.
\end{equation}
It contains all isolated roots of the original system which Bertini's 
theorem guarantees, see for example \S{1.1.4} in~\cite{Bates:2013}, and additional ``spurious'' roots
%
%
which do not solve (\ref{eq:FXC}) and which will differ, if $M$ is reshuffled. 
As shown later, at every step of the solution curve $X(S)=\{ E^+_k(S), E^-_k(S) \}$, see Fig. \ref{fig:pathtracker} (top left)
the residual of eq. (\ref{eq:FXCM}) will be decreasing. Simultaneously, at very step it
is tested if the residual of eq. (\ref{eq:FXC}) is decreasing (called ``down path'' later), the 
matrix $M$ reshuffled otherwise and the corresponding newly created reduced system (\ref{eq:FXCM}) solved,
colored path segments in Fig. \ref{fig:pathtracker} (top left). Then the solver cannot
converge to spurious roots corresponding to any particular $M$ along $X(S)$.


A standard iterative technique for root finding of nonlinear equations is Newton's method. 
Given an initial iterate $X_{n=0}$ 
\footnote{Abuse of notation, this is a variable vector of length $2k$. Here the index $n = 0$ denotes $0$th Newton iteration.}
and a nearby root $X^*$ the function (\ref{eq:FXCM}) is linearised at $X_0$
\begin{equation} \label{eq:F-linear}
   F^M(X_0) + F'^M(X_0)\Delta X - C_1^M = 0,
\end{equation}
solved for $\Delta X$, the Newton step and a step towards the root is taken
\begin{equation} \label{eq:newton-it}
   X_{n+1} = X_{n} + \Delta X.
\end{equation}
For a one-dimensional function $X_{n+1}$ is the point, where the tangent at $X_0$ crosses the $X$ axes. 
For vector valued functions the derivative $F'^M(X_0)$, the Jacobian, is a square matrix and 
eq. (\ref{eq:newton-it}) a linear system for the unknown $\Delta X$.
The iteration (\ref{eq:newton-it}) is known to converge roughly quadratically towards the root $e_{n+1} \sim e^2_{n}$,
if the function is Lipschitz continuous (which polynomial functions satisfy) and the Jacobian nonsingular, 
see for example {1.2.1} in \cite{Kelley:2003solving}. 
Where $e_n = \lVert  X^* - X_n \rVert$ is the error of the $n$th iteration with $\lVert \cdot \rVert$ being the standard 
Euclidean norm on $R^{2N}$. The roughly quadratic convergence can be observed monitoring the norm\footnotemark
$\lVert F(X_n) - C_1 \rVert$ often called \emph{residual}. 
\footnotetext{In the ultrafast optics community instead of the Euclidean norm, typically the rms error is used, often called FROG error or trace error.} 

For an arbitrarily chosen initial iterate $X_0$ there is, in general, no close enough root
for the iteration (\ref{eq:newton-it}) to converge, then, additional tricks are required to 
globalise Newton's method.
As the primary computational method for that purpose polynomial system solvers like Bertini and PHCpack 
\cite{Bates:2013,Verschelde:2011} employ the \emph{continuation} method, where a homotopy is assembled
\begin{equation} \label{eq:HXs}
   H^M(X,s) := \left(F^M(X) - C_0^M\right) (1-s) + \left(F^M(X) - C_1^M\right) s, \quad \text{with} \quad H^M(X(s=0),0) = 0 
\end{equation}
which is connecting two polynomial systems and all roots of them via smooth curves $X(s)$,
the start system (first term) at $s=0$ and the target system (second term) at $s=1$, where
$s$ is the continuation parameter; in general, a curve in the complex plane, in the 
following real $s\in[0,1]$.
 
An $X(s=0)$ is chosen freely (normally a Gaussian) to compute $C_0 := F(X(s=0))$ in forward direction.
It is guaranteed that beginning at the solution $X(s=0)$ of the so-created 
start system and following the curve $X(s)$ to arrive at a solution of the target system,
if $H'^M(X(s),s)$ is nonsingular and $s$ an arbitrary complex curve beginning at $s=0$ and ending at $s=1$.
Though, here we move along real paths where a finite number of singular
points\footnotemark exist, as an arbirary complex path would render the homotopy
to have complex coefficients and the continuation path would, in general, end
at an undesirable complex root of the target system. 

\footnotetext{The structure of singular points for real homotopies like (\ref{eq:HXs}) has been fully characterised
\cite{Li:1993}: These singularities are quadratic turning points or \emph{simple folds}, where two real and two complex
conjugated solution branches meet, rotated by $\pi/2$ in the complex plan and toughing at their turning points, the simple fold.
Both branches smoothly transit the turning point, if an arc-length parameter is used instead of $s$ 
or pseudo arclength continuation \cite{Keller:1987lectures}. Then it is possible to follow the real curve through the
bifurcation point or, alternatively, jump onto the complex solution branch. Following the 
real branch we simply return to a new real root of the start system, continuing the complex branch
we either end up on a complex root (or its complex conjugate) of the target system or eventually flow into 
another simple fold where a transition to another real branch is possible. We implemented pseudo arc-length
continuation. Unfortunately, after passing a simple fold along the complex branch it is unlikely 
that it touches another simple fold and rather ends up on an undesired complex solution of the
target system. The same phenomenon has been observed in \cite{Henderson:1990} in the attempt of
bypassing these singular points towards real roots.}

 
Starting at $s=s_m$ with $H^M(X(s_m),s_m) = 0$ and taking a step $s_{m+1} = s_m + \Delta s$ with 
$\Delta s$ small enough along the $X(s)$ curve, we can guarantee to be close enough 
to a solution of $H^M(X,s_{m+1}) = 0$ when using the initial iterate $X(s_m)$. This path tracking,
see Fig.~\ref{fig:pathtracker} (top left), is usually done in a predictor-corrector scheme with adaptive
step size control. We use step size parameters as in PHCpack \cite{Verschelde:2011}, a predictor given by the local tangent
and one Newton steps as a corrector (reusing the Jacobian to compute the new local tangent). 

%
For any particular $M$ we track the path $X(s)$, hold the tracker and reshuffel $M$ at a \emph{break point} $s=s_b$
\begin{equation}
   H^M(X,s_b) = F^M(X(s_b)) - C_b^M = 0, \quad \text{where} \quad C_b^M = C_0^M - s_b (C_0^M - C_1^M),
\end{equation}
whenever close to a singular point or if the residual of the full system 
(\ref{eq:FXC}) $\lVert \Delta C(s) \rVert = \lVert F(s) - C_1 \rVert$ increases, also called 
momentary distance to the target trace.
For the new randomly reduce system we use the intermediate solution $X(s_b)$ as
an initial iterate for the corresponding new homotopy beginning at $s=0$. In
this manner, we get a collection of path segments $\{ s_{b_i} \}_{i=1,\dots}$
where $S=s_{b_1} + s_{b_2} + \dots$ is the total continuation time, see Fig.
\ref{fig:pathtracker} (top left, colored segments), with decreasing $\lVert
\Delta C(S) \rVert$ (top right, colored).

The error of the reduced system (momentary distance to the reduced target trace) (top right, gray) $\lVert \Delta C^M(S) \rVert=$  
$\lVert F^M(S) - C_1^M \rVert = \lVert C_{b_i}^M - C_1^M \rVert$,
decreases linearly for each path segment 
\begin{eqnarray}
   \lVert C_{b_i}^M - C_1^M \rVert > \lVert C_{b_{i+1}}^M - C^M_1 \rVert = (1 - s_{b_i}) \lVert C_{b_i}^M - C_1^M \rVert \quad \Rightarrow \\
   \lVert \Delta C^M(S) \rVert \approx - \frac{d}{dS} \lVert \Delta C^M(S) \rVert \quad \Rightarrow \quad \lVert \Delta C^M(S) \rVert \approx \lVert \Delta C^M(S=0) \rVert e^{-S} 
\end{eqnarray}
and as the length of each path segment is relatively small $s_b \ll S$, globally, the total error decrease 
appears like an exponential decay in $S$. 

Initially, the distance $\lVert \Delta C(S) \rVert$ is decreasing approximately linearly as well on each segment.
We are moving along smooth curves, stepping along any newly create path with decreasing $\lVert \Delta C(S) \rVert$,
which we call \emph{down paths} as opposed to \emph{up paths}, it is likely that the next step is also decreasing.
The number of steps before reaching a break point ($\lVert \Delta C(S) \rVert$ increases) is getting smaller
as $X(S)$ is getting closer to the optimum, until no significant reduction
of the error is possible when reaching the accuracy limit set by numerical errors or noise floor. 

At every break point trial predictor-corrector steps are computed for newly created randomised systems until
a down path has been found. If the trail step succeeds (Newton's method converges), the path is 
called \emph{valid path}\footnotemark which can be either an up or down path.
\footnotetext{This automatically excludes all continuation paths for which the conditioning of the local Jacobian
is bad and those with high velocities / curvature. In practice, we first compute the full Jacobian and then try several 
random projections. This way we have the cost of computing the full Jacobian at the beginning of every new 
path segment only once.}
Clearly, the probability of finding an up path from the list of all valid paths is an important quantity 
which we denote as $p_\text{up}$\footnotemark. We found that almost every valid path is a down path before reaching 
the accuracy limit for noisy systems. Then, $p_\text{up}$ rises steeply, see Fig.~\ref{fig:pathtracker} (bottom right).   
This intrinsic quantity is, thus, most practical and sensitive to stop the solver by setting a threshold 
$\pUp > p_\text{up}$ at $\pUp = 50\% - 90\%$, for example.
\footnotetext{An efficient and still accurate enough method to estimate this quantity is, rather than computing many 
trail steps for every path segment, to keep a running list of the last, say, 20 trials of preceding path segments. 
The step size for computing trial steps for each path segment over the whole path has to be the same to make this
quantity comparable.}

\section{Adaptive regularizaton for noisy traces} \label{sec:regul}

In this section we show how to make the algorithm work for pulse retrieval of noisy experimental data. 

One effect of computing the integral (\ref{eq:one}) in the forward direction given $E$ is smoothing 
as we are dealing with an auto-correlation-like nonlinear integral.
Contrary, the inverse mapping acts as a high-pass filter with the undesirable tendency of noise amplification;
rather problematic when using Newton's method. We resolve this phenomenon by adding a regularization term to the equations. 
Tikhonov regularization or ridge regression, when solving an ill-posed least squares problem have a long history 
in statistics, see for example \cite{Engl:1996regularization}. 

The pixelwise smoothing of the trace (\ref{eq:JaveTot}) is providing noise reduction and thereby implicit regularization.
As the pixelization is refined, steep local gradients arise when sampling a set of continuation paths, causing the path tracker
to reduce the step size to very small and leading to smaller and smaller path segments. Until the smoothness
assumptions coming with Newton's method do not apply. Then more direct countermeasures are asked for.
\begin{figure}[!t]
\centering
\includegraphics[width=1.0\textwidth]{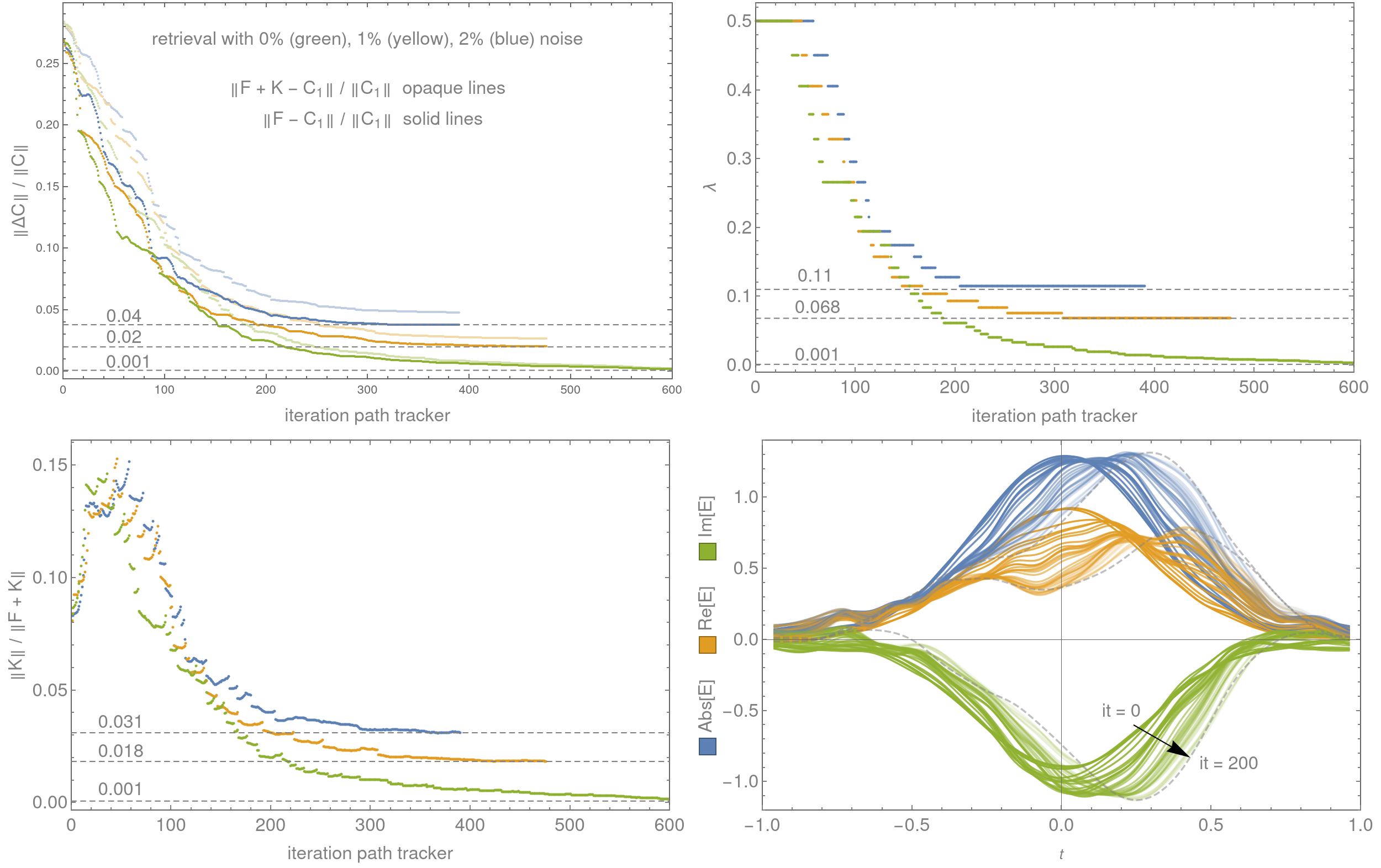}
\caption{\footnotesize Solving polynomial system including Tikhonov-type penalty term. Top right: The size of 
$\lambda$ is constant along each path segment (see Fig. \ref{fig:pathtracker}) and 
adaptively decreased at the end of each segment, if a threshold on the relative size of the regularisation term
is crossed $\delR < 20\%$ eq.(\ref{eq:delR}) for three different noise levels 0\% (green), 1\% (yellow), 2\% (blue). 
Bottom Left: The relative size of the penalty term $K$ on each segment is growing and globally adaptively reduced.
Top left: The distances to the target trace with (opaque colored) and without (full colored) penalty term are 
simultaneously decreasing along each path segment, while their difference is increasing because $F$ is deformed 
slightly towards an improved match with the target trace $C_1$ for a class of curves with similar mean curvature. 
Then, in the noiseless case (green) / noisy case (yellow, blue) $\delR$ vanishes / settles where $\lambda$ is
near optimal. 
Bottom right: Beginning at a zero phase Gaussian $E(s=0)$, connected by a series of smooth intermediate solutions
the path tracker continues towards the target pulse (dashed gray, only first 200 iterations shown).}
\label{fig:regul}
\end{figure}

Analogously to Tikhonov regularization we add a penalty term $K(X,\lambda)$ with components $K_l(X,\lambda),\, l=1,\dots, N(N+1)$ to 
the first $N(N+1)$ equations of the original system (\ref{eq:FXC}) which gives preference to solutions with smaller 
norms, also known as $L_2$ regularization
\footnote{For Tikhonov regularization the penalty term which is $\lVert X \rVert$ is added to the least squares
problem. Roots of the first derivative of this sum wrt $X$ correspond to regularized minima.} 
\begin{equation} \label{eq:FKC}
   F(X) + K(X,\lambda) - C_1 = 0, \quad K(X,\lambda) := \lambda\, M^i_{\text{reg}} \partial_i \lVert X \rVert^2 , 
\end{equation}
where $\lambda$ is the Tikhonov factor to scale the penalty term and $M^i_{\text{reg}}$ a shuffle matrix
which remains constant through out the path tracking and can be used for validation purposes via re-shuffling
and repeating the tracking.

For piecewise constant approximants (\ref{eq:0spline}) to $E(t)$ we get
\begin{eqnarray}
   \partial_l \lVert X \rVert^2 &=& \partial_l \sum^{2N-1}_{i=0} (X_{i+1} - X_i)^2 = 0, \quad l = 0,\dots,2N-1 \\
   &=& (X_{l-1} - 2 X_l + X_{l+1}) = 0,\, \text{and}\, (\text{optional boundary condition})
\end{eqnarray}
which is nothing but the 2nd order finite difference of $X$ on a three point stencil (up to a factor) which means
eq. (\ref{eq:FKC}) gives preference to solutions with small mean curvature and implements the desired smoothing effect.
An optional boundary condition can be added, for example $E^{+/-}_{k=-1} = 0$ and $E^{+/-}_{k=N} = 0$, where
we added two extra intervals on each end of the grid.

As the penalty term alters the solution as little regularization as necessary is wanted at the target solution,
though, while path tracking, $\lambda$ can be larger and this is actually beneficial from a numerical point of view
as it improves the conditioning of the Jacobian and smoother intermediate solutions enable longer path segments and larger steps.
By slowly decreasing\footnotemark $\lambda$, see Fig. \ref{fig:regul} (top right), we can assure that smooth initial data is 
connected by as series of smooth intermediate solutions to the smooth target Fig. \ref{fig:regul} (bottom right).
\footnotetext{On the other hand, throttling $\lambda$ too slowly during the path tracking, may cause an undesired prolongination of the path.}

Moving along any path segment and the related homotopy of (\ref{eq:FKC}), where we hold $\lambda$ constant, $F$ is deformed slightly towards
an improved match with the target trace $C_1$ for a class of curves with similar mean curvature. Therefore,
the difference $\rVert F + K - C_1 \lVert \,-\, \rVert F - C_1 \lVert$ is growing along each segment, while, of course,
both are decreasing simultaneously, see Fig. \ref{fig:regul} (top left). Until, to improve the match further, the amount of smoothness must 
be reduced by decreasing $\lambda$ (top right). Finally, the match cannot be significantly improved by allowing 
rougher curves. Even further reducing $\lambda$ would cause matching $F$ to the noise and the afore-mentioned
shortening of path segments and step size due to steeper local gradients; wasted computational cost.
The relative difference
\begin{equation} \label{eq:delR}
   \delR := \left( \rVert F + K - C_1 \lVert - \rVert F - C_1 \lVert \right)\, / \, \rVert F + K - C_1 \lVert
\end{equation}
is, thus, an ideal candidate for setting a threshold to lower $\lambda$ from one path segment to the other.
Moreover, $\delR$ is inert to details of the solution and noise model and vanishes, if no noise is present.  
For Fig. \ref{fig:regul} the threshold was set at $\delR < 20\%$.
 
When starting from very smooth initial data, like an initial Gaussian, in the coarse initial phase (first 100 iteration)
the size of the penalty term can grow undesirably Fig.~\ref{fig:regul} (bottom left) before the above mechanism can set in because
the intermediate solutions attain flection. (For the same reason the relative size of the penalty term
grows along each path segment.)
This is prevented by adding another threshold for decreasing 
$\lambda$, if $\rVert K \lVert \,/\, \rVert F + K \lVert$ rises above, say, $30\%$. Of course, if informed
initial data is at hand, like a solution from a coarser grid, this is not necessary.

As shown in the following section this adaptation mechanism is steering $\lambda$
near optimality or close enough for fine-tuning, for example, using the L-curve method or some other tool.

\section{Application examples, convergence, practical concerns} \label{sec:application}
 
We implemented the algorithm as a hybrid code in Mathematica (prototyping, pre- and post-processing) and in Fortran90 
(core routine path tracker). All simulations were performed on an Intel Core i7-4790 CPU\@3.6GHz with 4 cores on 8 Gb 
Ram, linux OS and using OpenMP parallelisation and the Intel Compiler. 

As a test cases we selected the pulse with index 42 (TBP2) from the data base of 101 randomly generated pulses with time-bandwidth
product (TBP) equal to 2 which were used in \cite{Geib:2019} to profile their least-squares solver, another less intricated 
test pulse (TBP1) using the same generator with TBP = 1 and a third test case (A2908).
For other test cases we found the same universal convergence behavior as shown in the following.
For every run a different seed is used to initialise the Xorshift random number generator ``xoshiro256+`` \cite{Vigna:2014} 
for computing the random matrices $M$. 

To show the applicability to realistic defective traces Gaussian noise is added with 
$\sigma_\text{noise} = 1\%,\, 2\%,\, 3\%$ to the synthetic trace 
$I_\text{exp}(\omega,\tau) = I[E_\text{in}](\omega,\tau) + noise$ before\footnotemark 
Fourier transforming it to $J_\text{exp}(\tau,\sigmaB)$.
\footnotetext{The noise is chosen to have zero mean, if this is not the case, either a background subtraction of $I_\text{exp}(\omega,\tau)$
has to be done or equivalently the zero mode after Fourier transform has to be removed and interpolated for $J_\text{exp}(\tau,\sigmaB)$ 
which we consider the cleaner choice.
The trace $J_\text{exp}(\tau,\sigmaB) \approx J[E_\text{in}](\tau,\sigmaB)$ is initially normalised to have its absolute value maximum equal to one. 
Then every initial data $E$ should be scaled for the integral $J[E](\tau,\sigmaB)$ to have the same property.}
We study the effect of varying, see eq. (\ref{eq:delR}), the regularisation reduction threshold $\delR = 5\%,\, 25\%,\, 40\%$ (controlling the decrease
of $\lambda$ while path tracking), see Sec. \ref{sec:regul}, as well as the effect of varying the termination criterion 
to stop the solver $\pUp = 50\%,\, 70\%,\, 90\%$ on the
error convergence while fine-graining the pixelization, increasing $N$. Where we measure the retrieval accuracy 
or \emph{pulse error} $\epsilon$ as
\begin{equation}
   \epsilon = \lVert E - E_\text{in} \rVert\, /\, \sqrt{N}\, /\, \text{max}(|E|) 
\end{equation} 
instead of using the relative error norm $ \lVert E - E_\text{in} \rVert / \lVert E \rVert$ to make results 
comparable with the literature, in particular, \cite{Geib:2019}. The scaling behavior with $N$ is the same for both metrics.
To measure the (pixelwise) \emph{trace error} we use the relative Euclidean distance to the target trace 
$\lVert \Delta C \rVert\, /\, \lVert C \rVert$ as before. In the literature on ultrafast nonlinear optics 
often the rms error, also called FROG error is used. 

\subsection{Initialisation, first example} \label{subsec:init}

As a first example, already discussed above, the retrieval of test pulse A2908 (dashed curves) is shown 
in Fig. \ref{fig:regul} (bottom right) for $N=25$ with 
$\sigma_\text{noise} = 1\%,\, \delR = 25\%,\, \pUp = 90\%$. Initial data was set to a Gaussian bell curve with zero phase,
where the initial width of the Gaussian is set to minimise the polynomial system including penalty term (\ref{eq:FKC})
for some large enough initial value of $\lambda$. 
 
The synthetic measurement trace with $129\times 129$ points was coarse grained to $N\times N = 25\times 25$ pixels.
In Fig.~\ref{fig:regul} (top left) the convergence of the trace error while iterating along the solution path is
shown, already discussed in a previous section.
When averaging over 100 retrievals we get a mean trace error / evolution time / within iterations: 
$0.04/1s/200, 0.02/2s/400, 0.019/2.5s/500$, compare with Fig. \ref{fig:regul} (top left). 
\begin{figure}[!t]
\centering
\includegraphics[width=0.6\textwidth]{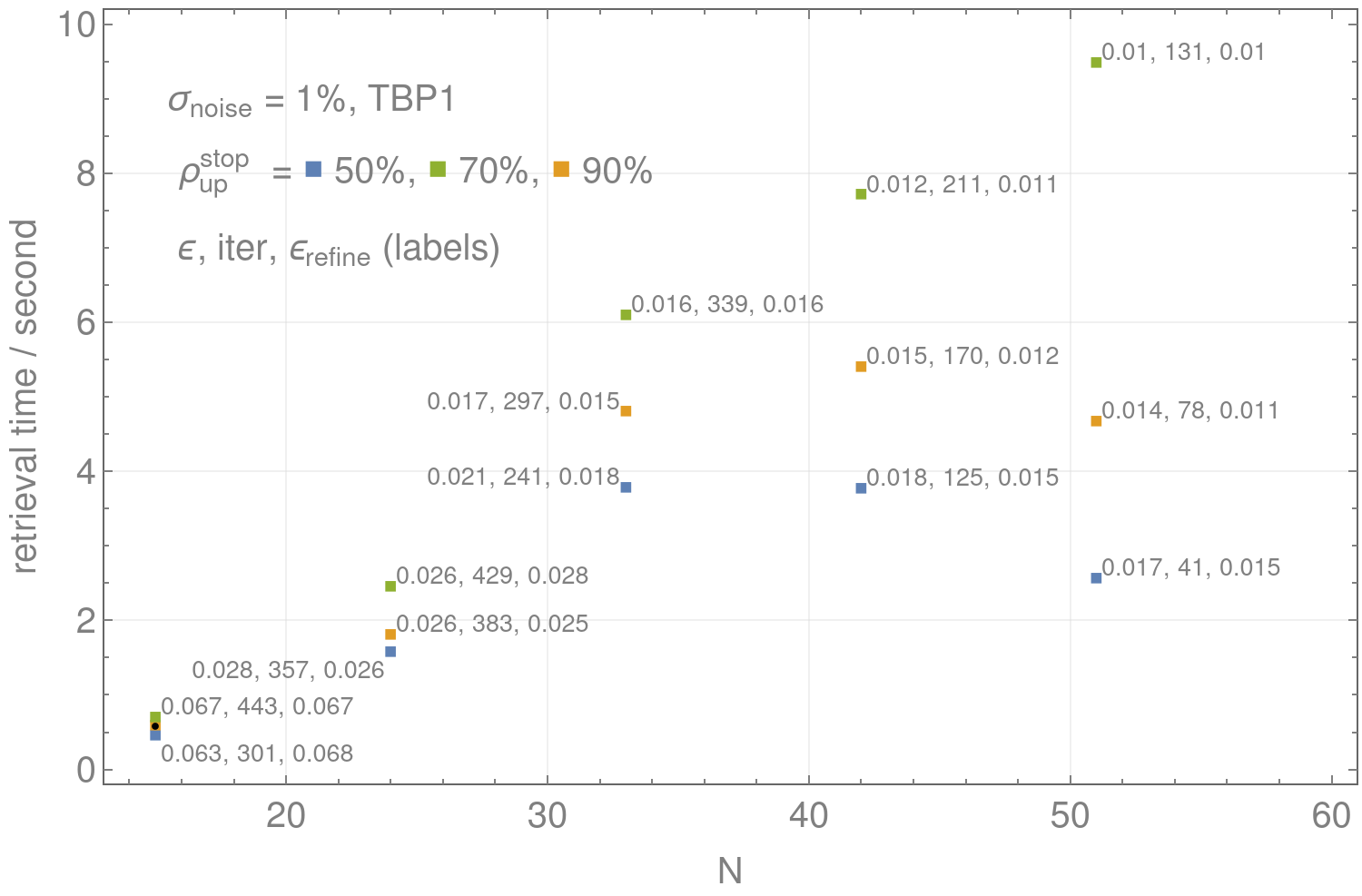}
\caption{\footnotesize Performance overview: Retrieval time vs $N$ with $1\%$ additive noise and three different values for the
termination criterion $\pUp$ and the resulting retrieval accuracy, number of iterations and
retrieval accuracy after applying additional refinement steps all averaged over 10 runs. Initial data for one level
is the interpolated result from the next coarser level. The plot implies to choose as smaller $\pUp$ while cascading
towards the final grid and do refined retrievals there.}
\label{fig:timing}
\end{figure}

For test pulses without gaps (points or regions with zero amplitude) like TPB1
and A2908 we found no cases of stagnation independent of the noise level, for
the retrieval probabilities with noise of the most common solvers see Fig. 7
\cite{Geib:2019}. For test pulse TBP2 about 2\% of all runs stagnate. In the
gap region $E^+$ and $E^-$ are restricted to be zero and configurations\footnotemark appear
\footnotetext{Taking as an example a double pulse (two pulses separated by a
gap) with quadratic phase we found that two false configurations appear which
have a small trace error and the correct quadratic phase along each of the two
pulses but a phase jump of $+\pi$ or $-\pi$ at the gap. For larger gaps we
found a greater likelihood to run into these configurations in numerical
experiments.}
that cannot be freely deformed into one another under the constraint of moving
along real ``down paths''. Introducing a lifted electrical field by adding a
large enough constant $E \rightarrow E + c$ as a new variable which has by
construction no gap resolves the problem which we study in more detail in \cite{Jasiulek:2021}. 

At the moment the following practical workaround can be used: first,  
compute $k$ trial runs ($k\approx5$) with a zero phase initial Gaussian, zero phase on 
a coarse grid $N=15$ to $N=25$ which takes about $0.3s$ to $1s$ per run. 
Then, the correct initialisation for a fine grid retrieval using this result as initial data has chances $1-0.10^k$,
if the coarse grid retrievals fail in, for example, $10\%$ of the cases.

\subsection{Retrieval timing, computational cost} \label{subsec:timing}

\begin{figure}[!t]
\centering
\includegraphics[width=0.95\textwidth]{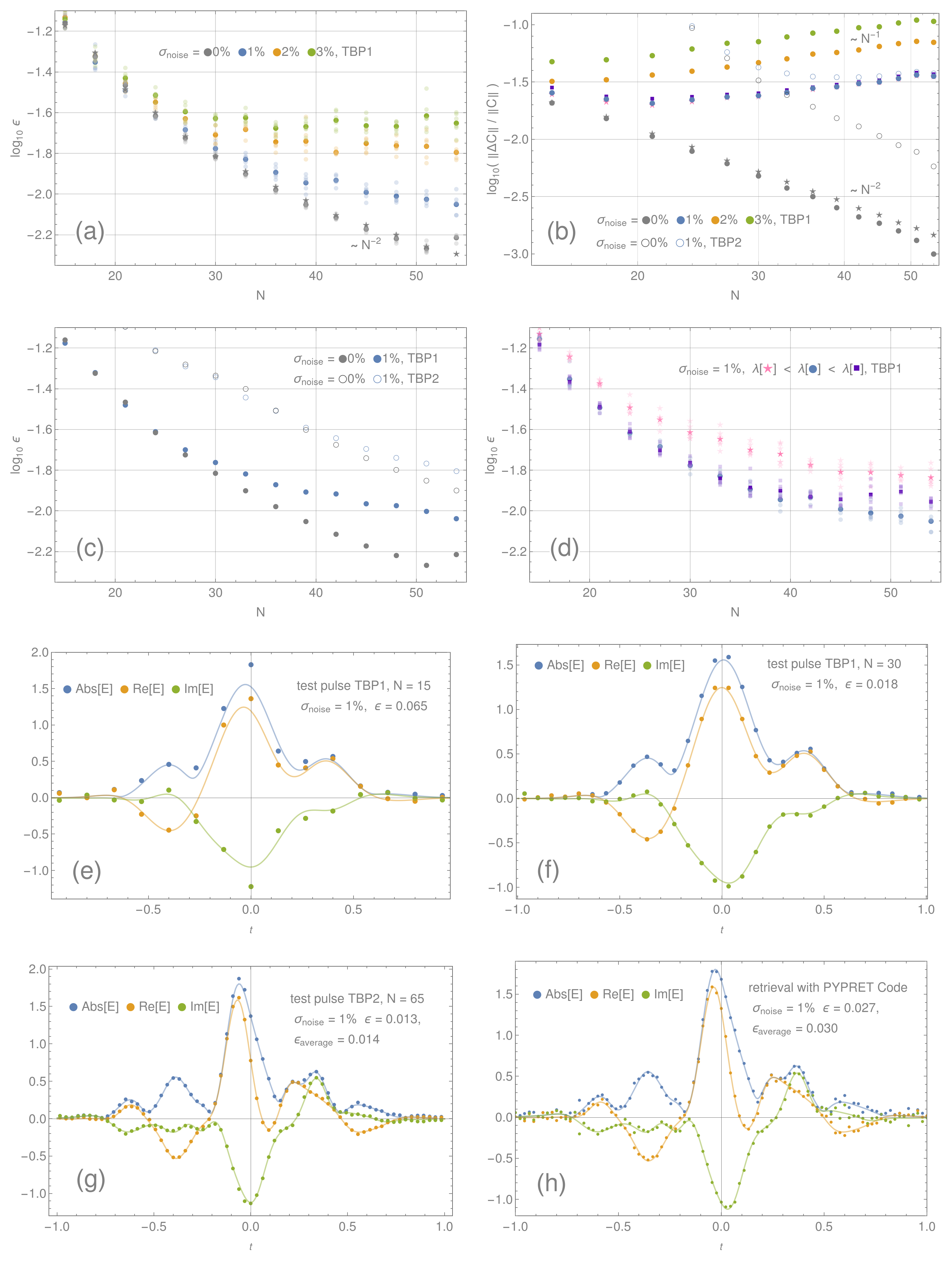}
\caption{\scriptsize Error convergence vs $N$: Average (full colored points) final pulse error $\epsilon$ (a,c,d) and trace error 
$\lVert \Delta C \rVert / \lVert C \rVert$ (b) on 14 pixelisations for the test cases TBP1, TBP2 with and 
without noise.
Two examples of the retrieved electric field on a coarse (e) and medium (f) pixelised grid for TBP1.
(g) vs (h): Comparison of our solver with the result of a least-squares solver without regularisation.
For more details see Sec. \ref{sec:convergence}. Note: Here more intermediate grids then necessary are
used. Normally, a cascade like $N \rightarrow 20 \rightarrow 40 \rightarrow 65$ is sufficient.
}
\label{fig:convergence}
\end{figure}

A more detailed overview of the timing, number of iterations and the influence of the termination criterion
of the implemented algorithm is shown in Fig.~\ref{fig:timing}. Every data point corresponds to an average over
10 runs. As a test case we used the pulse TBP1 with $\sigma_\text{noise} = 1\%$. 
Initial data for one level is the interpolated result from the next coarser level. Then, the number of iterations 
before termination decreases approximately linearly with $N$ for noisy traces and is constant for noiseless cases
(not shown here). As mentioned before, a speedup by a factor of two is possible, as any of the summands in 
eq. (\ref{eq:JaveTot}) suffice to approximated the total pixel average. Here we use both terms. 
 
A larger value of $\pUp$ results in
higher accuracy but the additional costs usually do not justify the small improvement on $\epsilon$
at every intermediate level. Fig.~\ref{fig:timing} suggests using $\pUp = 50\%$ on coarser grids before reaching
the target level and doing refinement there, if required. If speed is a concern, the number of intermediate
levels can be optimised. For our purposes a single initial coarse grid and one fine grid was enough.
As an example $N_\text{initial} = 15,\, N_\text{target} = 51$, we got 
$\epsilon_\text{target} \approx 0.015,\, \text{iterations} \approx 600,\, \text{retrieval time} \approx 30 s$. 

Another practical concern, if the investigated pulse has large low-amplitude wings, a large part of the computational 
domain and cost are spend on this low amplitude region and a clipping or zoom of the experimental data to
the region of interest is suggestive, setting the wings to zero at first. Then, in a follow-up retrieval a 
larger domain could be included using the result as initial data, if required.

\subsection{Convergence, scaling behavior, comparison} \label{sec:convergence}

In Fig.~\ref{fig:convergence} (a-d) the convergence of pulse and trace error are studied 
on 14 grid levels $N = 15,18,\dots,53$ with 6 retrievals per $N$ (opaque colored points) 
and their averages (full colored points) for test pulses TBP1 and TBP2. 
In (e-g) the input pulse shapes are shown (solid lines) and a retrieval result (points) on $N=15$ (e), $N=30$ (f), $N=51$ (g).
We set $\pUp = 90\%$ and do an additional centering and two refinement steps at every level, see Subsec.\ref{subsec:refine}. 

The retrieval accuracy $\epsilon$ without noise (gray dots and stars) Fig.~\ref{fig:convergence} (a) as 
well as the trace error converge with $\sim 1/N^2$ (gray dots and stars) Fig.~\ref{fig:convergence} (b) 
as the dominant numerical error is stemming from the interpolating of $J[E](\tau,\sigmaB)$ from boundary values
to the pixel interior (\ref{eq:lowerLeft}) which is of the order $O(h^2)$.
This scaling behavior is overlayed with a $\sim N$ increase of the (pixelwise) relative
trace error for noisy traces $J_\text{exp}$, as the noise suppression scales with $\sim 1/\sqrt{N^2}$ when averaging over 
less data points per pixel area for larger $N$.
Still $\epsilon$ may decrease while $\lVert \Delta C \rVert$ is increasing as more details of the pulse 
are being resolved when increase $N$, until 
$\epsilon$ becomes approximately constant at the accuracy limit, see Fig.~\ref{fig:convergence} (c), 
for TBP1 with 1\% noise (solid circles) at about $\epsilon = 10^{-2},\, N=50$ for TBP2 with 1\% (empty circles) 
at about $\epsilon = 10^{-1.9},\, N=65$. 
Resolving more details beyond this point is possible, if less noisy traces are input. The dependence of 
the accuracy limit on the noise level is apparent in Fig.~\ref{fig:convergence} (a) (blue, yellow, green). 
Analogously, for larger enough $N$ the minimal trace error should only depend on the amount of noise and not the
particular pulse. This is apparent in Fig.~\ref{fig:convergence} (b) (empty / filled gray and blue points). 

To make the numerical integration error apparent when computing pixel surface averages from the trace 
$J_\text{exp}(\tau,\sigma) \rightarrow \langle J^\text{exp}_{ij} \rangle$ we chose two different resolutions
$129\times 129$ points (gray dots) vs $257\times 257$ (gray stars) in Fig.~\ref{fig:convergence} (a,b).
The integration error is only relevant, if the noise level is low enough to reach this high accuracy.
\footnotetext{We use the trapezoidal rule for numerical integration and first order interpolation, if the
pixel boundaries do not automatically lie on data grid. First order interpolation was used to preserve the
additive nature of the noise. For realistic traces, where other sources of error dominate, higher order 
interpolation could be used.}


An interesting effect becomes apparent when varying the amount of regularisation by varying the threshold 
$\delR = 5\%, 25\%, 40\%$ (pink, violet, blue), zoom 
into Fig. \ref{fig:convergence} (b). Though, the trace error is larger for larger $\lambda$ (violet, blue vs pink) the corresponding 
pulse error, as shown in (d), is smaller (does not apply to even larger $\lambda$). 
This is a reminder that retrieval techniques / optimisation codes that only aim at minimising the trace error without regularisation  
are prone to fitting the noise; the right balance between over-fitting and over-smoothing has to be found. 
Consider this when comparing a retrieval of our algorithm with the result of the least-squares 
solver (no regularisation\footnote{An unpublished version (private repository) of this solver including 
regularisation is available by now.}) used in \cite{Geib:2019}, see Fig. \ref{fig:convergence} (g) vs (h) \footnotemark.
\footnotetext{More grid points (parameters) not necessarily imply higher resolution and accuracy. 
An increase beyond the number of significant parameters (effective DOF) for least-squares can 
cause overfitting. In Fig. \ref{fig:convergence} (g) for TBP2 with 1\% noise the pulse error did improve beyond 
$N=65,\, \epsilon = 0.013$ and similar to TBP1 with 1\%, see Fig. \ref{fig:convergence} (c), the accuracy limit is in both 
cases at about $10^{-2}$. As the position of the pulse relative to the grid is not fixed, the result can be sampled on 
arbitrary inter-grid locations, see next section.}
As this effect on the trace error is relatively small, one could also express it differently: there are many possible 
pulse shapes of varying smoothness which have approximately the same trace error but a rather different pulse error. 
Through regularisation and coarse-graining an optimal pulse shape can be computed.

Due to Tikhonov regularization the presented solver provides a smoother and
more accurate solution in the pulse tails as well as a smaller total retrieval
error\footnotemark, see Fig. \ref{fig:convergence} (g) vs (h), or Fig.
\ref{fig:lcurve} (right) vs Fig. \ref{fig:convergence} (h), in comparison with
the least-squares solver of \cite{Geib:2019}; the computational time compares
to $\approx 30s$ (or $15s$, see Footnote \ref{footnote:speed}) vs $\approx
10s$. This solver had the lowest retrieval error for noisy data in comparison
with the most common existing solvers, see Fig. 7 in \cite{Geib:2019}.  
Another advantage of the presented solver is that the retrieval probability does
not depend on the noise level and that the cases of stagnation can be fully resolved,
see SubSec. \ref{subsec:init} and \cite{Jasiulek:2021}.  
\footnotetext{The relatively larger error of the presented solver at the pulse's peak near $t=0$ 
disappears when refining the resolution.}
%

\subsection{Optimal $\lambda$, refined solution, oversampled solution} \label{subsec:refine}

\begin{figure}[!t]
\centering
\includegraphics[width=1.05\textwidth]{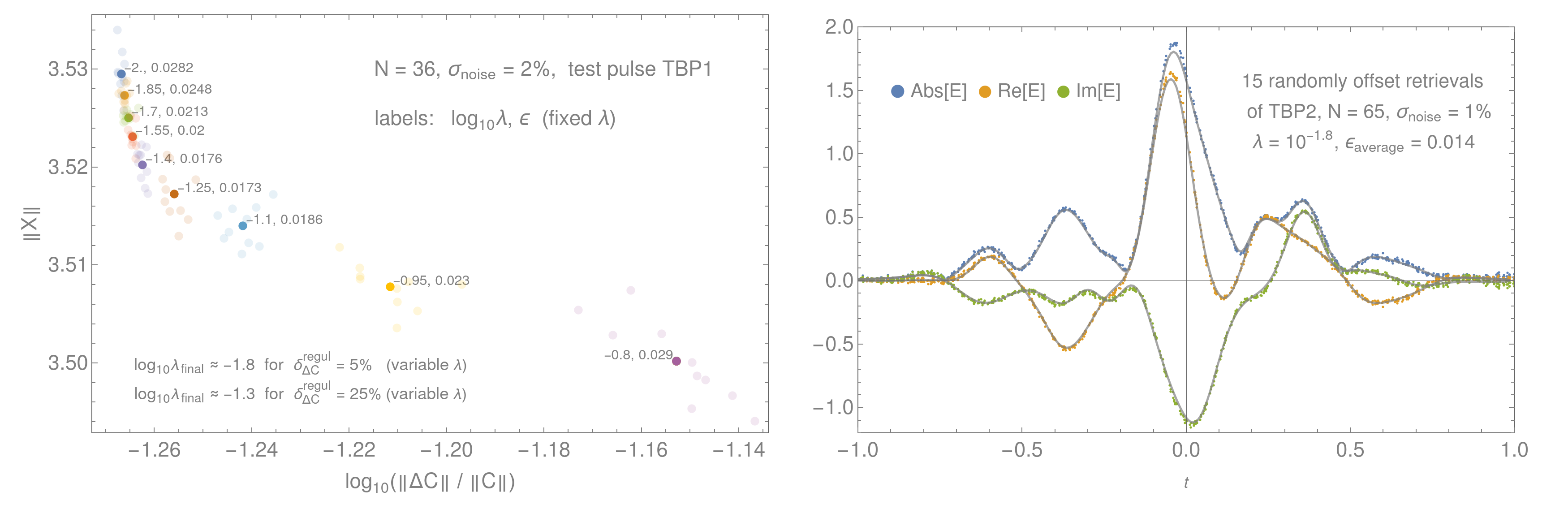}
\caption{\footnotesize Left: Applying the L-curve method for 9 different values of $\lambda$, 10 retrievals each (opaque colored),
mean value (full colored). Noisy oscillations begin to increase the length of the pulse $\lVert X \rVert$ when decreasing $\lambda$,
the trace error $\lVert \Delta C \rVert$ improves through over-fitting. Too smooth solutions (too large $\lambda$) show deviations 
from the original solution and have larger trace errors. The optimal amount of $\lambda$ is within the corner of the L-curve where 
the pulse error $\epsilon$ is small.
Right: 15 randomly offset retrievals of pulse TBP2 (gray lines) on $N=65$ intervals. As the position of the pulse 
is not fixed relative to the numerical grid, sampling on arbitrary inter-grid locations is possible.}
\label{fig:lcurve}
\end{figure}

The most natural choice to test and fine tune the amount of regularisation for optimality in the presents of additive noise 
would be a chi-square test while varying $\lambda$ because the pulse error is not available a priori. 
%
Considering modern measurement devices and pulse retrieval setups, see for example \cite{Geib:2020}, for the problem at hand, 
where (difficult to quantify) systematical errors and multiplicative noise are the most relevant sources of error, a goodness 
of fit test of this type does not seem applicable yet to a measured trace. 
A popular practical solution that we recommend in this case, is the so-called \emph{L-curve} method \cite{Hansen:1999} until more
sophisticated techniques are asked for.

An application of the L-curve method for test case TBP1 with $\sigma_\text{noise} = 2\%,\, N = 36$ is shown in Fig. \ref{fig:lcurve},
where for fixed $\lambda$ (adaptive decrease of $\lambda$ turned off) 10 retrievals (opaque colored points) and the average 
(full colored points) are shown. For smallest $\lambda$ the retrieved pulses have smallest trace error (overfitting region) 
but they do not have the smallest pulse error. The length $\lVert X \rVert$ of each pulse is extended by noisy wiggling
about some smoother solution which is reached for increasing $\lambda$.
In the L's corner, the amount of regularisation is optimal and the pulse error is minimal.
 
Note that the overfitted solutions also reveal themselves by having the smallest spread (opaque colored points) in the 
trace error as they differ only by a new random sample of the reverse amplified noise; all samples having in average 
about the same length. 
 
The implemented mechanism to adaptively decrease $\lambda$ while path tracking, Sec. \ref{sec:regul}, steers
$\lambda$ towards optimality. Here are two examples: $\log_{10} \lambda_\text{final} = -1.8$ for $\delR = 5\%$ and for 
$\delR = 25\%$ $\log_{10} \lambda_\text{final} = -1.3$ which is optimal.
 
As the ratio of up paths relative to all valid paths $p_\text{up}$ is estimated from a finite sample, the number
of iterations before crossing the threshold set by $\pUp$ differ slightly. To assure the solver cannot improve
the solution within this threshold we perturb the solution sightly by shifting it one (or more) grid points to the
left or right and use it as initial data for a new retrieval. This refinement step does not improve $\epsilon$
much, if the threshold was already set high $\pUp \approx 90\%$, as shown in Sec. \ref{subsec:timing}, Fig. \ref{fig:timing},
but yields a small improvement otherwise. 

There is a translational symmetry which has not been discussed yet: $E(t) \rightarrow E(t+\delta t)$. This symmetry is, 
strictly speaking, broken as the $E(t)$ is defined on a bounded domain $t\in[-1,1]$ and zero elsewhere. But as we are
dealing with finite accuracy solutions and as $E(t)$ models a physical light 
pulse with low amplitude
wings there are actually infinitely many similar solutions within a given error bound which differ only by a small 
shift $\delta t$ of the pulse relative to the numerical grid. As a consequence, in particular on coarser grids any solution
should be centered on the numerical grid before transferring the result.
As a benefit, on finer grids, if we re-process a solution shifted by some small random inter-grid distance $\delta t < h$ 
via interpolation, the result is a shifted solution, sampled on slightly different points $E(t+\delta t)$.
With this technique the pulse can be sampled on arbitrary inter-grid locations, as shown 
in Fig. \ref{fig:lcurve} (right), where the above procedure was applied for $N=65$ for TBP2 with 1\% noise.  
To be more precise, ``the'' over-sampled solution is rather an error band as for noisy traces a spread of 
near optimal solutions within some error bound exist.  
A small deviation from the original pulse is apparent when looking at its peak-value which is caused by the 
discretisation error of $E(t)$ on $N=65$ intervals that should disappear as the grid is refined.
We did the same as above for the pulse TBP2 but with $\lambda = 10^{-1.7}$ (not shown). The average pulse error and 
oscillations in the error band were slightly less, though, small deviations of the mean curve through 
the error band from the original pulse are visible in some regions. Implying that care has to be taken when fine-tuning 
$\lambda$ at the corner of the L-curve; rather choosing solutions with slightly less $\lambda$, slightly
bigger $\lVert X \rVert$ and averaging to obtain a mean curve. This phenomenon is also apparent in 
Fig. \ref{fig:lcurve} (left, compare green, red, violet, brown) all four points have small nearby errors, 
the smallest for brown, though, not the optimal choice.

\section{Conclusion and Outlook} \label{sec:conclusion}

An algorithm has been developed having in mind the applicability to real experimental data 
for common pulse retrieval schemes in ultrafast nonlinear optics like FROG, d-scan or
amplitude-swing \cite{Trebino:1993,Miranda:2012,Alonso:2020}. 
The employed numerical techniques have been successfully applied in other fields of physics, where nonlinear integral 
equations of similar type appear, techniques used in polynomial system solvers and stochastic optimisation.
It has been shown how to implement Tikhonov-type regularisation into the polynomial equations, how to 
adaptively decrease it while path tracking the solution and how to fine-tune $\lambda$ at the 
solution when dealing with noisy, defective experimental data.
The integral equation was discretized to a polynomial system such that each equation corresponds 
to a grid cell (pixel) surface average of the original integral equation, rather than a pointwise representation. 
This coarsening capability enables fast computations of approximants, noise suppression 
and high accuracy retrievals on fine pixelizations.

The full solution path is a collection of small linked continuation paths, along each the error decreases 
monotonically by construction, where these path segments are associated to different random projections of 
the full polynomials system.
This constitutes a new method for path tracking real solutions of similar over-determined polynomial systems
through these partially continuous and stochastic solution paths. 
Each movement along a single path segment causes a deformation of the momentary solution in parts in the 
direction of the global solution and in parts towards some random perturbation. These perturbations 
appear to cancel when path tracking over a collection of many path segments.
Similarly, perturbations due to added noise seem to compensate each other and an optimal solution can be computed.
A link to the theoretical framework developed for the Newton-sketch method \cite{Pilanci:2017newton,Berahas:2020}, 
or in Randomised Numerical Linear Algebra \cite{Drineas2016:randnla,Martinsson:2020} or in other areas of 
stochastic optimisation seems plausible. The main difference is that the continuation method is implemented
here. This may also be advantageous when computing the global minimum of non-convex optimisation problems.

If speed is a concern, there are two ways to accelerate the solver: The most immediate solution
is parallelising the list auto-correlations in (\ref{eq:lowerLeft}) on GPUs. 
Secondly, it stands to reason trying Jacobian-free Newton-Krylov methods \cite{Knoll:2004jacobian} 
or other Quasi-Newton methods \cite{Kelley:1995iterative,Kelley:2003solving} to replace Newton's method 
in the algorithm which could accelerate the computations by a factor of $N$. 
Thirdly, when reducing the full system to random linear combinations the Hadamard transform
is the method of choice for dimensional reduction \cite{Ailon:2009,Boutsidis:2013} for many other applications in
stochastic optimisation. 

For realistic FROG or d-scan traces additional frequency dependent functions can be introduced
to the nonlinear integral which model frequency dependent systematical errors in the nonlinear medium
and the experimental setup which are otherwise neglected. These functions could
be added to the list of unknowns and retrieved with the presented algorithm, similar to \cite{Miranda:2017}.


For refining a retrieved pulse further the pointwise representation of the integral equation could be 
used as well, see eq.\ (\ref{eq:list-auto}), if the quality of the measured data admits this. Then the 
numerical integration errors could be avoided with a small gain in accuracy. 

Finally, there are other similar phase retrieval problems in science and engineering \cite{Candes:2015,Fannjiang:2020} 
and in optics and nonlinear optics \cite{Fienup:82,Walther:1963,Shechtman:2015,Mairesse:2005}, where an 
application of this algorithm seems plausible. 

\section*{Acknowledgement}

I would like to thank Günter Steinmeyer, Esmerando Escoto, Lorenz von Grafenstein from the MBI Berlin,
Peter Staudt from APE GmbH, as well as Carl T. Kelley, Michael H. Henderson, Jan Verschelde,  Alex Townsend
for helpful comments and discussion.
In particular, I thank Nils C. Geib for carefully reading and commenting on the first version of this article.
%
%
This work was financially support by the European Union through the EFRE program 
(IBB Berlin, ProFit grant, contract no. 10164801, project OptoScope) in 
collaboration between the Max-Born Institute, Berlin and the company APE GmbH.

\bibliographystyle{unsrt}
\bibliography{newton}

\appendix
\section{Appendix} \label{app:A}

The algorithm is applicable to other similar pulse retrieval schemes when replacing 
the nonlinearity in the integral (\ref{eq:shg_frog}),  
$E(t)E(t-\tau) \rightarrow E(t)H(t-\tau)$. For example, for polarisation-gate 
FROG or third-harmonic-generation FROG we have 
\begin{equation}
   H(t) = E(t) \EB(t),\, E(t)^2 \nonumber 
\end{equation}
and the integral (\ref{eq:Jshg}) becomes 
\begin{eqnarray}
 J[E,\bar{E}](\tau,\sigmaB) = \int_{-\infty}^{+\infty} E(t) H(t-\tau) \bar{E}(t-\sigmaB) \bar{H}(t-\tau-\sigmaB) \,\text{d}t, \nonumber \\ 
 F_\tau(t) := E(t)H(t-\tau),\quad G_\sigmaB(t) := E(t)\bar{H}(t-\sigmaB). \quad \quad \quad \nonumber
\end{eqnarray}
For the pulse retrieval scheme d-scan \cite{Miranda:2012}, 
again applying the convolution theorem, we get
\begin{eqnarray}
 I[E](z,\omega) = \left\lvert \int_{-\infty}^{+\infty} F_z(t)^2 e^{-i \omega t}  \,\text{d}t \right\rvert^2 \quad &\rightarrow& \quad
 J[E](z,\sigmaB)  = \int_{-\infty}^{+\infty} F_z(t)^2 \bar{F}_z(t-\sigmaB)^2 \,\text{d}t, \,\, \nonumber \\ 
 F_z(t) := \int_{-\infty}^{+\infty} \hat{E}(\omega) \hat{\phi}(\omega,z) e^{i \omega t} \,\text{d}\omega \quad &\rightarrow& \quad 
 F_z(t)  = \int_{-\infty}^{+\infty} E(t) \phi(s-t,z) \,\text{d}s,
\end{eqnarray}
where the function $\hat{\phi}(\omega,z): = e^{i\, z\, k(\omega)}$ \footnotemark is known and given
by $z$ the thickness of the material in the beam path, $k(\omega)$ the dispersion of 
the material, $\hat{E}(\omega)$ and $\hat{\phi}(\omega,z)$ are the Fourier transforms of $E(t)$ and $\phi(t,z)$.
\footnotetext{For the pulse retrieval scheme amplitude-swing $\hat{\phi}:=
\hat{\phi}(\omega,\theta)$ is a function of $\omega$ and the angle $\theta$, the relative orientation of
a rotating birefringent material and a linear polarizer \cite{Alonso:2020}.}
As before, see eq. (\ref{eq:Jauto}), along lines of constant $z$ the integral $J[E](z,\sigmaB)$ is a one-dimensional
auto-correlation of the function $F_z(t)^2$ and the pixel average can be computed with (\ref{eq:corrLeftRight}), 
(\ref{eq:JaveTot}). To compute $F_z(t)$ an additional complex convolution
with the material function $\phi(s,z)$ has to be evaluated instead of a multiplication. When using piecewise-constant
approximants for $E(t)$ and $\phi(s,z)$ the corresponding list convolution can be computed as before 
with eq. (\ref{eq:list-auto}). 


Another useful formula: an equivalent form of the nonlinear integral (\ref{eq:one}) 
using the frequency domain representation $\hat{E}(\omega)$ of the enveloping electric field is
\begin{equation} 
    I[E](\omega,\tau) := \left\lvert \int_{-\infty}^{+\infty} \hat{E}(\omega-\Omega) \hat{E}(\Omega) e^{i\Omega \tau} \,\text{d}\Omega \right\rvert^2,
\end{equation}
applying a Fourier transform as before but in the variable $\tau \rightarrow \rho$ and with an opposite sign,
the frequency domain representation of the integral $\hat{J}[E](\rho,\omega)$ appears to be an auto-convolution 
instead of an auto-correlation in the new auxiliary variable $\hat{F}_\omega(\Omega)$ 
\begin{equation} 
 \hat{J}[E](\rho,\omega) = \int_{-\infty}^{+\infty} \hat{F}_\omega(\Omega) \bar{\hat{F}}_\omega(\rho-\Omega) \,\text{d}\Omega, \quad 
 \hat{F}_\omega(\Omega) := \hat{E}(\Omega) \hat{E}(\omega-\Omega) 
\end{equation}
Therefore, instead of retrieving $E(t)$ the whole formalism can be rephrased to retrieve $\hat{E}(\omega)$.
The same applies to the other pulse retrieval schemes from above.

\section{Appendix} \label{app:B}

It is possible to use polynomials of higher order, like splines in each interval, compare with eq. (\ref{eq:0spline})
\begin{equation} \label{eq:1spline}
    E(t) =  \nonumber 
    \begin{cases} 
          0   & t < -1 \quad \text{or} \quad 1 < t \\
          \sum^p_{A=0}E^A_k\,\th^A & t \in [t_k,t_{k+1}], \quad k = 0,\dots, N-1 
    \end{cases}
\end{equation}
with polynomial order $p$ and $\hat{t}$ is again the local interval coordinate. 
Then the function $F_\tau(t)$ 
\begin{eqnarray}
   ^{(i)}F_k(\hat{t}) &=& E^0_k E^0_{k+i} + ( E^0_k E^1_{k+i} + E^0_k E^1_{k+i} )\hat{t} + \dots \nonumber \\
   ^{(i)}F_k(\hat{t}) &=& {^{(i)} C^0_k} + {^{(i)} C^1_k}\, \hat{t} + \dots + {^{(i)} C^{2p}_k}\, \hat{t}^{2p} \nonumber \\ 
   ^{(i)}\FB_k(\hat{t}) &=& {^{(i)} D^0_k} + {^{(i)} D^1_k}\, \hat{t} + \dots + {^{(i)} D^{2p}_k}\, \hat{t}^{2p} \nonumber 
\end{eqnarray}
discretises into a polynomial of order $2p$ for which the integration in eq. (\ref{eq:list-auto}) becomes 
\begin{eqnarray} \label{eq:Jspline} 
   J[E](\tau_i,\sigmaB) = h \sum^{(2p+1)^2}_{m=0} \int_{\sigmaBH-1}^1 \hat{t}^{A(m)} (\hat{t}-\sigmaBH)^{B(m)} d\hat{t} \;
                        \sum_{k=1}^N\,^{(i)}C^{A(m)}_k\, ^{(i)} D^{B(m)}_{k+j} + \dots, 
\end{eqnarray}
where we use an index vector $(A(m),B(m))$ in degree lexicographical order with lowest order first. 
\begin{eqnarray} \label{eq:corrLeftRight}
 \langle J[E]_{ij}^\text{left right} \rangle = 
   \frac{h}{4} \sum^{(2p+1)^2}_{m=0} v_m  \left( \text{corr}( {^{(i  )} C^{A(m)}}_k, {^{(i  )} D^{B(m)}}_{k} )_j     + \text{corr}( {^{(i+1)}C^{A(m)}}_k, {^{(i+1)} D^{B(m)}}_{k} )_{j} \right) \nonumber\\
+\; \frac{h}{4} \sum^{(2p+1)^2}_{m=0} w_m \left( \text{corr}( {^{(i  )} C^{A(m)}}_k, {^{(i  )} D^{B(m)}}_{k} )_{j+1} + \text{corr}( {^{(i+1)}C^{A(m)}}_k, {^{(i+1)} D^{B(m)}}_{k} )_{j+1} \right) \nonumber
\end{eqnarray}
As an example, consider a polynomial chain $p=1$, then $^{(i)} C^A_k,\, A=0,\dots,2$ and 
$(2p+1)^2 = 9$ list correlations that have to be computed on every interval as compared to 1 for $p=0$.
The index vector is $(A(m), B(m)) = ( (0,0), (0,1), (1,0), (0,2), (1,1), (2,0), (1,2), (2,1), (2,2) )$. 
The canonical integrals over $\hat{t}$ in eq. (\ref{eq:Jspline}) turn into a list of integers when integrating over $\hat{t}$ and $\sigmaBH$ 
\begin{eqnarray}
   v_m &=& (2,\; -2/3,\; +2/3,\; 2/3,\; 0,\; 2/3,\; +2/15,\; -2/15,\; 2/9)  \nonumber \\ 
   w_m &=& (2,\; +2/3,\; -2/3,\; 2/3,\; 0,\; 2/3,\; -2/15,\; +2/15,\; 2/9). \nonumber
\end{eqnarray}

\end{document}